\documentclass[aps,11pt,final,notitlepage,oneside,onecolumn,nobibnotes,nofootinbib,superscriptaddress,noshowpacs,centertags]{revtex4}

\usepackage{amsmath,amssymb,epsfig,placeins,subfigure,wrapfig,indentfirst}

\begin{document}

\title{A new method for reconstructing the density distribution of matter\\ 
in the disks of spiral galaxies from the rotation velocity curve in it}

\author{Alexander Shatskiy}
\affiliation{Astro Space Center, Lebedev Physical Institute, Russian Academy of Sciences, Moscow, Russia}

\author{Igor Novikov}
\affiliation{Astro Space Center, Lebedev Physical Institute, Russian Academy of Sciences, Moscow, Russia}
\affiliation{Niels Bohr Institute, Copenhagen, Denmark}
\affiliation{Kurchatov Institute,  Moscow, Russia}

\author{Olga K. Sil'chenko} 
\affiliation{Sternberg Astronomical Institute, Moscow State University, 13 Universitetski prospect, 119992, Moscow, Russia}
\affiliation{Isaac Newton Institute of Chile, Moscow Branch}

\author{Jakob Hansen}
\affiliation{Korea Institute of Science and Technology Information (KISTI), 335 Gwahak-ro, Yuseong-gu, Daejeon, 305-806, Korea}

\author{Ivan Yu. Katkov} 
\affiliation{Sternberg Astronomical Institute, Moscow State University, 13 Universitetski prospect, 119992, Moscow, Russia}

\date{\today}

\begin{abstract}

{\bf ABSTRACT}

In this paper we propose a new method for reconstructing the surface density of matter in flat disks of spiral galaxies.
The surface density is expressed through observational rotation velocity curves of visible matter in the disks of spiral galaxies.
The new method is not based on quadrature of special functions. 
The found solution is used for processing and analysis of observational data from several spiral galaxies.
The new method can be used to more accurately estimate the amount of dark matter in spiral galaxies. 

\end{abstract}

\maketitle

\section{Introduction}
\label{s1}

Let us consider a model of a spiral galaxy with all matter distributed in a thin disk (thickness disk is neglected).
Accounting for the components of a spherical halo of the galaxy see in section~\ref{s6}. 

Let us consider the axisymmetric model in cylindrical coordinates ${(\rho, z, \varphi)}$.  

In the disk we assume a stationary rotation of dust-like matter, i.e. equality (modulo) of the gravitational and centrifugal forces:
\begin{eqnarray}
\frac{\partial\Phi(\rho , z=0)}{\partial\rho} = \frac{V^2(\rho)}{\rho} 
\label{1-1}\end{eqnarray}
Here ${V(\rho)}$ is the rotation velocity curve of the matter and ${\Phi(\rho , z)}$ is the Newtonian gravitational potential of the dust matter, which may be expressed as: 
\begin{eqnarray}
\Phi(\rho , z)=-G\int\limits_0^\infty R\, dR\, \int\limits_{-\pi}^\pi d\varphi \, 
\frac{\sigma (R)}{\sqrt{\rho^2+R^2-2\rho R\cos\varphi +z^2}}
\label{1-3}\end{eqnarray} 
where ${\sigma (\rho)}$ is the distribution of the surface density of matter in the disk. 

We pose the problem: Express the distribution of the matter surface density, ${\sigma (\rho)}$, in the disk from the rotation velocity curve of the matter ${V(\rho)}$. 

The potential ${\Phi(\rho , z)}$ is a function of ${\sigma (\rho)}$. 
Gauss's theorem for this potential leads to the expression: 
\begin{eqnarray}
\left. \Phi(\rho , z),_z\right|_{z=0} = 2\pi G\sigma(\rho)  
\label{1-2}\end{eqnarray}
Here $G$ is the gravitational constant. 

Thus, we can obtain an integral equation expressing the velocity $V$ by the integral of the surface
density ${\sigma}$. 
This equation can be inverted (see~\cite{Toomre1963, Kostov2006}) to express the surface density ${\sigma}$ through the rotation velocity curve in the disc: 
\begin{eqnarray}
\sigma (\rho )=\int\limits_0^\infty dk\, J_0(k\rho) \tilde\sigma(k)\, ,\quad 
\tilde\sigma(k)\equiv \frac{k}{2\pi G} \int\limits_0^\infty dR\, J_1(kR)\, V^2(R) 
\label{2-1}\end{eqnarray}
Here $J_0(x)$ and $J_1(x)$ are Bessel functions. 

Using this method to determine the surface density ${\sigma(\rho)}$ has significant shortcomings: 

1. To find the desired function ${\sigma(\rho)}$ it is necessary to integrate a fourfold integral, whereas the Bessel functions themselves are each expressed in terms of integrals. 

2. In the integrand of (\ref{2-1}) are oscillating and slowly decaying (as ${\cos x/\sqrt{x}}$) Bessel function, which also makes it difficult to integrate.

3. The integrals has infinite upper limits and therefore they requires a large numerical resources (change of variables leads to new difficulties due to the oscillating integrands).
 
All of these deficiencies substantially affect the quality of the results (obtained by numerical methods).

In papers~\cite{Jalocha2008, Bratek2008, Bratek2010}, a more convenient method based on integrals of elliptic functions. 
It was shown that the main errors arise due to ignorance of the distribution rotation velocity curve ${V(\rho)}$ at large values of ${\rho}$. 
The same problem (stability of the numerical calculation of $\sigma$) was studied in~\cite{Polyachenko2004, Polyachenko2005}.

In this paper we propose a new method for solving the problem of determining the surface density ${\sigma(\rho)}$ from the disc velocity curve ${V(\rho)}$.  
Our method does not require knowledge of the numerical values of special functions (Bessel functions and/or elliptic functions).

\section{The new method} 
\label{s2} 

Using the relation ${J_1(x)=-J_0(x),_x}$ for the Bessel functions and applying integration by parts to the integral for ${\tilde\sigma(k)}$ we rewrite expression (\ref{2-1}): 
\begin{eqnarray}
2\pi G\sigma (\rho )=\int\limits_0^\infty dk \int\limits_0^\infty dR\,\, J_0(k\rho) J_0(kR)\, [V^2(R)],_R  
\label{2-2}\end{eqnarray}
Here we took into account that 
$$
-\int\limits_0^\infty dR\, \frac{1}{k} J_0(kR),_R\, V^2(R) =
-\left. \frac{1}{k} J_0(kR)\, V^2(R)\right|_0^\infty + 
\frac{1}{k}\int\limits_0^\infty dR\,\, J_0(kR)\, [V^2(R)],_R 
$$
Also we took into account that at ${x=0}$ the function $J_0(x)$ is limited and $V^2(x)$ vanishes (no velocity on the axis). At ${x\to\infty}$ the opposite is the case, i.e. the function $V^2(x)$ is limited and the function $J_0(x)$ decays as 
${\cos(x-\pi/4)\sqrt{2/(\pi x)}}$. 

We denote the gradient of the square of the velocity as $f(R)$: 
\begin{eqnarray}
f(R)\equiv [V^2(R)],_R  
\label{2-3}\end{eqnarray}
Substituting the integral definition of Bessel functions of zero order   
\begin{eqnarray}
J_0(x)=\frac{1}{\pi} \int\limits_0^{\pi} e^{ix\cos\alpha}\, d\alpha    
\label{2-4}\end{eqnarray}
into eq. (\ref{2-2}), gives: 
\begin{eqnarray}
2\pi G\sigma (\rho )=\frac{1}{2\pi^2}\int\limits_0^{\pi} d\alpha \int\limits_0^{\pi} d\beta \int\limits_{-\infty}^\infty dk 
\int\limits_0^\infty dR\,\, e^{ik\rho\cos\alpha} e^{ikR\cos\beta}\, f(R)  
\label{2-5}\end{eqnarray}

It was here taken into account that $J_0(x)$ is an even dunction, so that the integration over $k$ in (\ref{2-2}) can be made by changing the limits of integration to go from ${-\infty}$ to $\infty$ and multiplying the result with a factor ${\frac{1}{2}}$.

Let us now consider the integral representation of the Dirac delta-function: 
$$
\delta (x)=\frac{1}{2\pi} \int\limits_{-\infty}^\infty e^{ikx} \, dk  
$$
and substitute it in (\ref{2-5}): 
\begin{eqnarray}
2\pi G\sigma (\rho)= \frac{1}{\pi}\int\limits_0^{\pi} d\alpha \int\limits_0^{\pi} d\beta  
\int\limits_0^\infty dR\,\, \delta (\rho\cos\alpha + R\cos\beta)\, f(R) 
\label{2-6}\end{eqnarray}
Here we take into account the integration rules of the complex delta-function: 
$$
\delta [g(R)]=\frac{\delta(R-R_0)}{|g(R),_R|}
$$
where $R_0$ are the simple roots of the equation ${g(R_0)=0}$. Also we should have that ${R_0\ge 0}$, which means that the integration region with respect to $\alpha$ and $\beta$ reduces to the two sectors: 
\begin{eqnarray}
1.\left\{ 
\begin{array}{rcl}
\alpha\in [0;\,\pi/2] \\ 
\beta\in [\pi/2;\, \pi] \\ 
\end{array} 
\right. 
\qquad 2. \left\{ 
\begin{array}{rcl}
\alpha\in [\pi/2;\,\pi] \\ 
\beta\in [0;\, \pi/2] \\ 
\end{array} 
\right.
\label{2-7}\end{eqnarray}
By renaming the variables $\alpha$ and $\beta$ in these sectors\footnote{In the first case, we redefine ${\alpha=\tilde\alpha}$, ${\beta=\pi/2+\tilde\beta}$; in the second case: ${\alpha=\pi-\tilde\alpha}$, ${\beta=\pi/2-\tilde\beta}$.} we may reduce the integral (\ref{2-6}) to the form: 
\begin{eqnarray}
2\pi G\sigma (\rho)= \frac{2}{\pi}\int\limits_0^{\pi/2} d\alpha \int\limits_0^{\pi/2} d\beta\,\, 
\frac{f(R_0)}{\sin\beta}\, , 
\quad R_0=\rho\frac{\cos\alpha}{\sin\beta}  
\label{2-8}\end{eqnarray} 
This expression, (\ref{2-8}), is the desired result which is without the deficiencies mentioned in section~\ref{s1}. 

Note that in expression (\ref{2-8}) we used the so-called holographic principle, i.e. each point of the density distribution ${\sigma(\rho)}$ contains the information about all points of distribution ${f(R)}$. 
This is due to integration over the angles $\alpha$ and $\beta$. Integration over the angle $\alpha$ provides a scanning range ${[0,\,\rho]}$, and integration over the angle $\beta$ provides a scanning range ${[\rho ,\,\infty)}$. 
In addition, ${\sin\beta}$ in the denominator of the integrand provides increased weight of the integrand ${f(R)}$ at large $R$. 
The converse is also true; each point of the distribution ${f(R)}$ contains information about all points of the density distribution ${\sigma(\rho)}$ - see section~\ref{s4}.  

This interesting feature is characteristic for the mass distribution in a flat disk\footnote{In the case of spherical symmetry, the model will no longer possess such a property; each point $R$ in the distribution of velocity will carry information only about the total mass inside a sphere with a radius $R$, and value of the mass density will be associated only with the pressure of matter (not with velocity).}. 

Using the formula ${V^2(R_0),_\rho = V^2(R_0),_{R_0}\cdot R_0,_\rho = f(R_0) \cos\alpha /\sin\beta}$ we can rewrite eq. 
(\ref{2-8}) in the form: 
\begin{eqnarray}
2\pi G\sigma (\rho)=\frac{d U(\rho)}{d\rho}\, ,\quad
U(\rho)\equiv\frac{2}{\pi}\int\limits_0^{\pi/2} d\alpha
\int\limits_0^{\pi/2} d\beta\,\, \frac{V^2(R_0)}{\cos\alpha}\,
,\quad U(\rho)= \int\limits_0^\rho 2\pi G\sigma (\rho)\, d\rho . \label{2-9}\end{eqnarray}
Using this expression it is convenient to obtain an expression for the total mass of matter inside the disk radius $\rho$ by integrating eq. (\ref{2-9}) by parts: 
\begin{eqnarray}
GM(\rho)=\int\limits_0^\rho 2\pi G\rho\sigma (\rho)\, d\rho = U(\rho)\cdot \rho - \int\limits_0^\rho U(\rho)\, d\rho  
\label{2-10}\end{eqnarray}

\section{Checking the method}
\label{s4}

\begin{figure*}
\includegraphics[width=15cm]{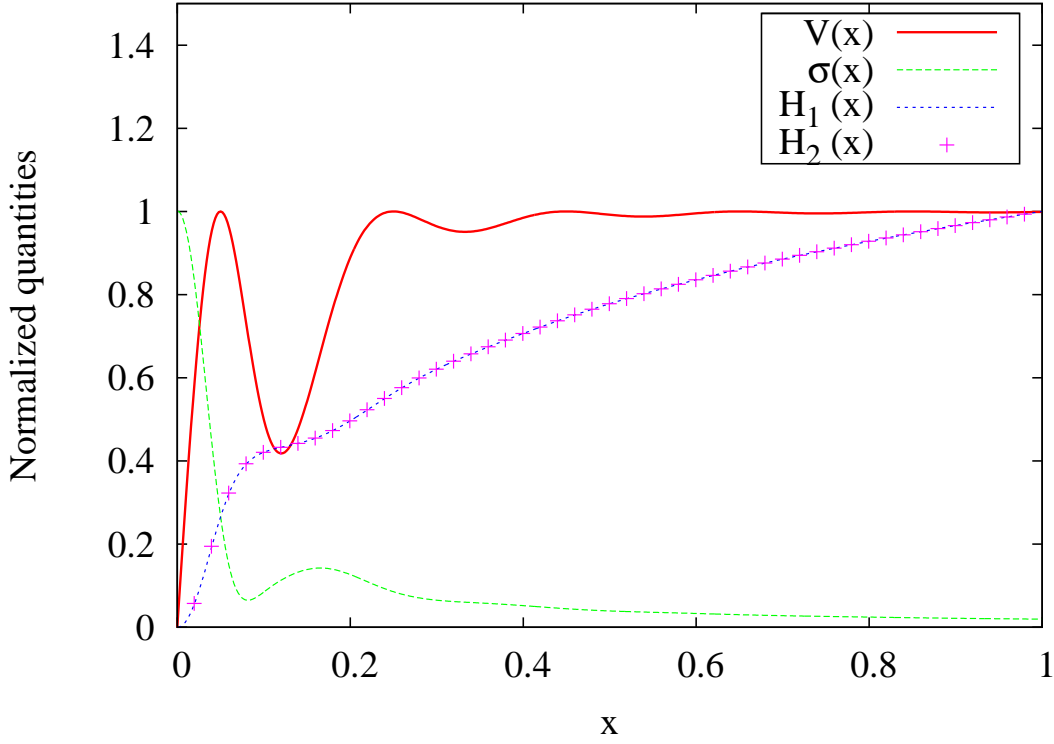} 
\caption{{Curves: ${V(x)=\frac{(10x)^3+\sin (10\pi x)}{1+(10x)^3}}$ - red line, ${\sigma(x)/\sigma(0)}$ - 
green line, ${H_1(x)}$ - blue line, ${H_2(x)}$ - red crosses (overlapping with ${H_1(x)}$)}; ${x\equiv R/R_{max}}$.}
\label{R1}
\end{figure*}

For basic testing of our new method, we use the following approach;
We integrate the expression~(\ref{1-1}) with respect to $\rho$: 
\begin{eqnarray}
H_1 (\rho)\equiv \int\limits_0^{\rho} \frac{V^2(\rho)}{\rho}\, d\rho 
\label{4-1}\end{eqnarray}
On the other hand, this same expression can be obtained from (\ref{1-3}):
\begin{eqnarray}
H_2(\rho)=U(\infty)-\frac{2}{\pi}\int\limits_0^\infty dR \cdot 2\pi
G\sigma (R)\, \int\limits_{0}^{\pi/2}d\gamma\,\,
\frac{R}{\sqrt{(R-\rho)^2+4R\rho \sin^2\gamma}} \, ,\quad
\gamma\equiv \varphi /2 \label{4-2}\end{eqnarray}
Or as:  
\begin{eqnarray}
H_2(\rho)=\frac{2}{\pi}\int\limits_0^\infty dR \cdot 2\pi G\sigma
(R)\, \int\limits_{0}^{\pi/2}d\gamma\,
\left(1-\frac{R}{\sqrt{(R-\rho)^2+4R\rho \sin^2\gamma}}\right)
\label{4-3}\end{eqnarray}
The integrals (\ref{4-2}) and (\ref{4-3}) do not have non-integrable
singularities and can therefore easily be found numerically. 

Expressions for the functions ${H_1(\rho)}$ and ${H_2(\rho)}$ ought to be
identical, as they must be equal to the expression  
${\Phi(\rho ,\, z=0)-\Phi(0,\, z=0)}$. 
Thus, we may judge the practical precision of our new method by evaluating and
comparing the expressions for ${H_1(\rho)}$ and ${H_2(\rho)}$ (using our new
method to calculate $\sigma(R)$).

To evaluate and compare these expressions, and thus to put our new method to
use, we have written a simple numerical code, the details of which are described
in Appendix A. In Fig. ~\ref{R1} is shown an example of such a comparison, where
we have used the sample velocity function :
\begin{eqnarray}
{V(x)=\frac{(10x)^3+\sin (10\pi x)}{1+(10x)^3}}
\label{velocity-test-func}\end{eqnarray}
as the basis for data. 
In the figure this velocity function and $\sigma (x)$ are calculated by using our new method. 
Also, of course, is plotted the two functions  ${H_1(x)}$ and ${H_2(x)}$ that we
wish to compare.

As can be seen from the figure (and as has been confirmed by detailed analysis of the
figure), the two functions ${H_1(x)}$ and
${H_2(x)}$ agree very closely. This is also the case for other velocity
functions we have tested. Thus we conclude that our new method is both valid and
accurate, even when doing the integrations numerically.

\section{Special cases and asymptotic distribution of the density of matter}
\label{s3}

In some special cases, expression (\ref{2-8}) can be reduced to an ordinary quadrature and we may find the asymptotic behaviour of ${\sigma(\rho)}$. 

1. Consider the simplest case for the distribution function ${f(R)}$ and the rate ${V(R)}$: 
\begin{eqnarray}
f(R)= \left\{ 
\begin{array}{rcl}
f_0 \makebox{\quad for  } R\in [0;\, R_1] \\ 
0 \makebox{\quad for  } R\in (R_1;\, \infty) \\ 
\end{array} 
\right. 
\qquad  
V(R)= \left\{ 
\begin{array}{rcl}
\sqrt{f_0\cdot R} \makebox{\quad for  } R\in [0;\, R_1] \\ 
\sqrt{f_0\cdot R_1} \makebox{\quad for  } R\in (R_1;\, \infty) \\ 
\end{array} 
\right. 
\label{3-1}\end{eqnarray}
The integral over ${\beta}$ in expression (\ref{2-8}) can be easily evaluated and we may obtain the quadrature:  
\begin{eqnarray}
2\pi G\sigma (\rho)= \frac{2 f_0}{\pi}\int\limits_{\alpha_1}^{\pi/2} 
\ln\left[ \frac{1+\sqrt{1-(\rho \cos\alpha /R_1)^2}}{\rho \cos\alpha /R_1} \right] d\alpha , 
\,\, \alpha_1= \left\{ 
\begin{array}{rcl}
0 \makebox{\quad for  } \rho \le R_1 \\ 
\arccos (R_1 /\rho) \makebox{\quad for  } \rho > R_1 \\ 
\end{array} 
\right. 
\label{3-2}\end{eqnarray}
or in another form: 
\begin{eqnarray}
2\pi G\sigma (\rho)= \frac{2R_1 f_0}{\pi\rho}\int\limits_{0}^{x_1} 
\frac{\ln\left[ (1+\sqrt{1-x^2})/x \right]}{\sqrt{1-(x R_1 /\rho)^2}}  \, dx \, , 
\quad  x_1= \left\{ 
\begin{array}{rcl}
\rho /R_1 \makebox{\quad for  } \rho \le R_1 \\ 
1 \makebox{\quad for  } \rho > R_1 \\ 
\end{array} 
\right. 
\label{3-3}\end{eqnarray}
From expression (\ref{3-2}) we find that the asymptotic behaiour at ${\rho /R_1\to 0}$ goes as ${\sigma(\rho)\propto \ln (R_1/\rho)}$. \\
From expression (\ref{3-3}) we find that the asymptotic behaviour at ${\rho /R_1\to \infty}$ goes as ${\sigma(\rho)\propto (R_1/\rho)}$. \\
It is noted that the asymptotic behaviour of $ {\sigma (\rho)} $ has a weak singularity at the origin. This is connected with the non-rigid rotation of matter around the central axis (for the case of rigid rotation, we should have ${V(R)\propto R}$ at ${R\to 0}$) - see eq.~(\ref{3-1}).

\begin{figure*}
\caption{{Observed and calculated data for the galaxies:\\ 
${(a)\, NGC2841}$, ${(b)\, NGC7217}$, ${(c)\, NGC7331}$ and ${(d)\, NGC5533}$.\\    
The horizontal axis is the distance in units of $x$  ${(x\equiv R/R_{max})}$.\quad On the vertical axis plots:\\    
${A)\, -\, V(x)/V(1)}$;\\       
${B)\, -\, M(x)/M_{max}}$; 
${\qquad C)\, -\, \sigma(x)/\sigma_{max}\quad - }$ for ${V(R>R_{max})=V(R_{max})=const}$; \\   
${B')\, -\, M(x)/M_{max}}$; 
${\qquad C')\, -\, \sigma(x)/\sigma_{max}\quad - }$ for ${V(R>R_{max})=V(R_{max})\sqrt{R_{max}/R}}$; \\
${D)\, -\,}$ vertical line marking the end of the stellar disk to the 25-th blue isophotes.}}
\subfigure[\label{2841} NGC2841: ${R_{max}=63\, 240 Pc}$, \qquad
${A)\,\, V(1)\approx 294 km/sec}$,  \qquad   
${B)\,\, M_{max}=M(1)\approx 1.22\cdot 10^{12} M_\odot}$, 
${C)\,\, \sigma_{max}=\sigma(0)\approx 3\, 432 M_\odot /Pc^2}$, 
${C')\,\, \sigma(0)\approx 3\, 406 M_\odot /Pc^2}$. 
]{\includegraphics[width=0.45\textwidth,  height=0.30\textheight]{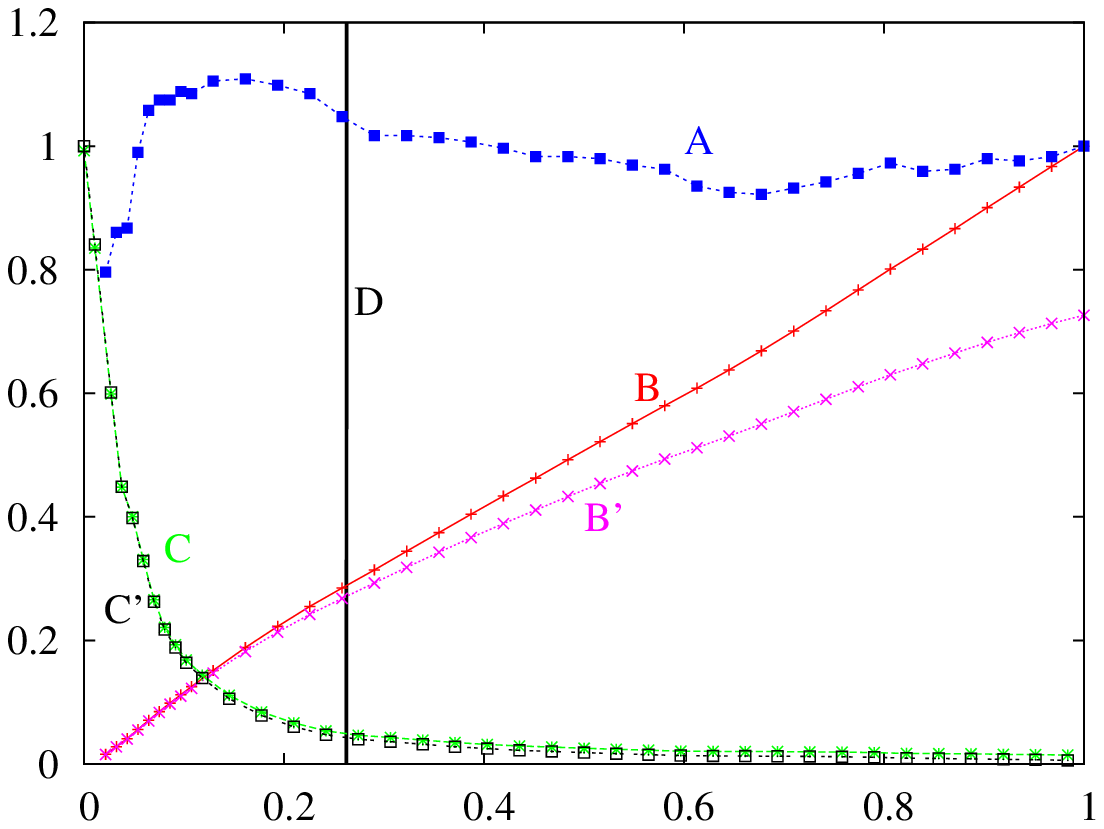}} 
\subfigure[\label{7217} NGC7217: ${R_{max}=9\, 600 Pc}$, \qquad
${A)\,\, V(1)\approx 305 km/sec}$,  \qquad   
${B)\,\, M_{max}=M(1)\approx 1.94\cdot 10^{11} M_\odot}$, 
${C)\,\, \sigma_{max}=\sigma(0)\approx 6\, 220 M_\odot /Pc^2}$, 
${C')\,\, \sigma(0)\approx 6\, 050 M_\odot /Pc^2}$.  
]{\includegraphics[width=0.45\textwidth, height=0.30\textheight]{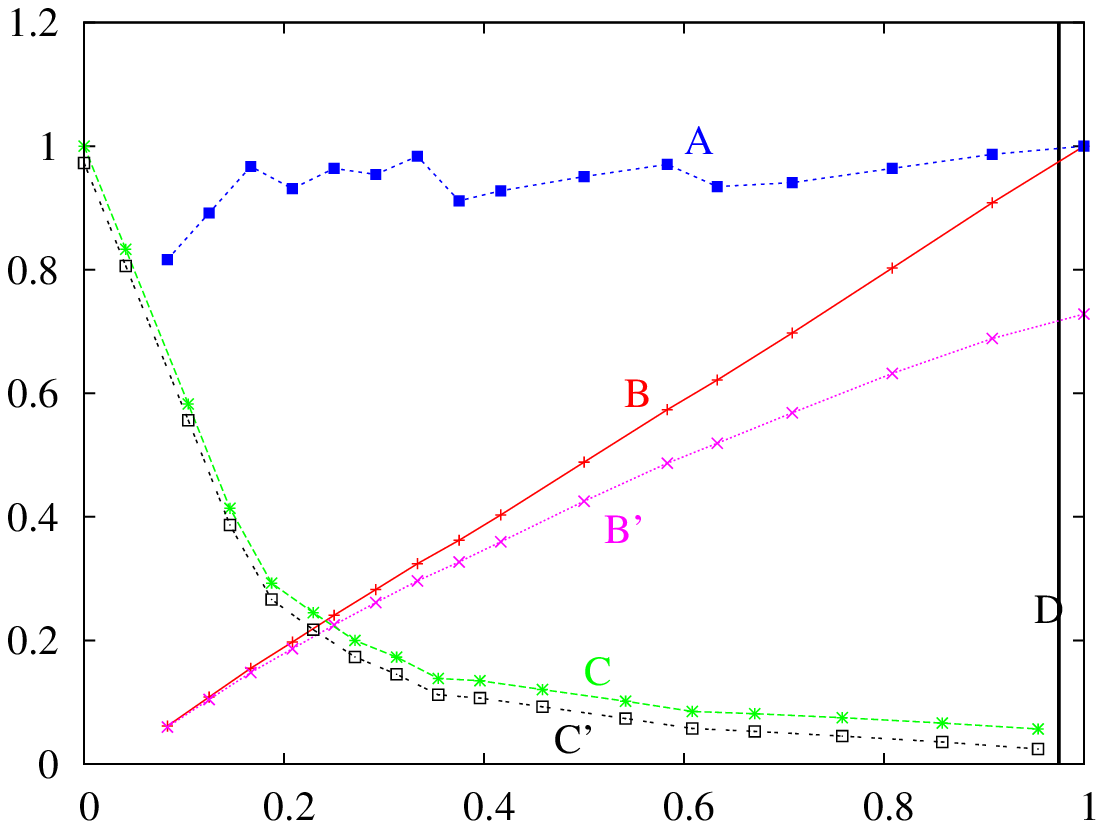}} 
\subfigure[\label{7331} NGC7331: ${R_{max}=37\, 230 Pc}$, \qquad
${A)\,\, V(1)\approx 238 km/sec}$,  \qquad  
${B)\,\, M_{max}=M(1)\approx 4.93\cdot 10^{11} M_\odot}$, 
${C)\,\, \sigma_{max}=\sigma(0)\approx 1\, 883 M_\odot /Pc^2}$, 
${C')\,\, \sigma(0)\approx 1\, 854 M_\odot /Pc^2}$.  
]{\includegraphics[width=0.45\textwidth, height=0.30\textheight]{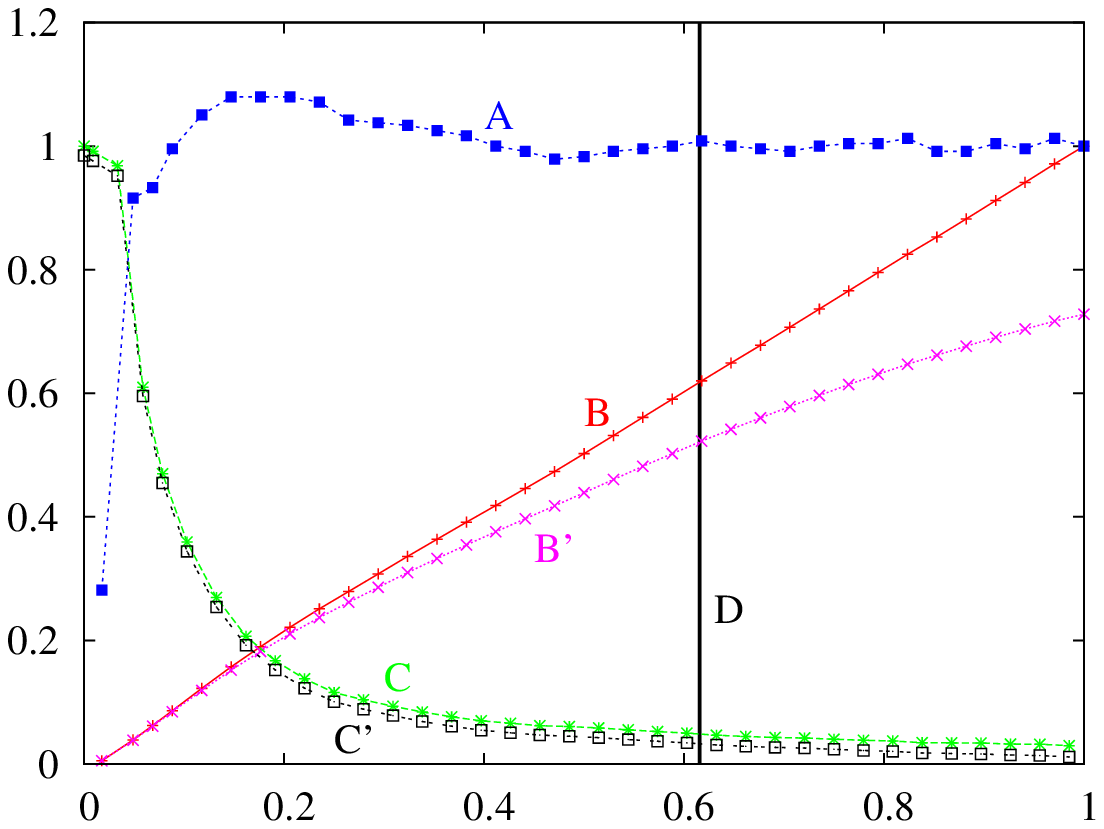}} 
\subfigure[\label{5533} NGC5533: ${R_{max}=77\, 875.2 Pc}$, \qquad
${A)\,\, V(1)\approx 226 km/sec}$, \qquad  
${B)\,\, M_{max}=M(1)\approx 9.34\cdot 10^{11} M_\odot}$, 
${C)\,\, \sigma_{max}=\sigma(0)\approx 2\, 877 M_\odot /Pc^2}$, 
${C')\,\, \sigma(0)\approx 2\, 867 M_\odot /Pc^2}$.  
]{\includegraphics[width=0.45\textwidth, height=0.30\textheight]{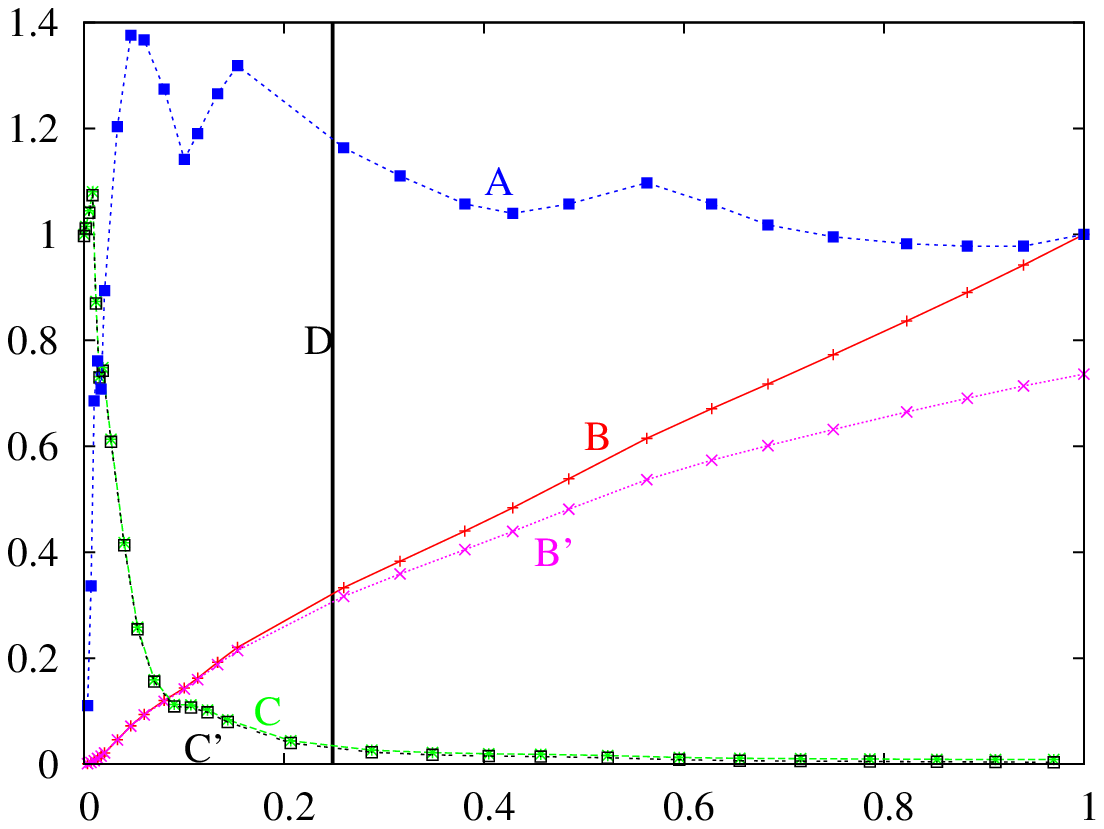}} 
\label{R2}\end{figure*}

2. We now consider another simple case, that of bounded rigid-body rotation: 
\begin{eqnarray}
f(R)= \left\{ 
\begin{array}{rcl}
f_0\cdot R/R_1 \makebox{\quad for  } R\in [0;\, R_1] \\ 
0 \makebox{\quad for  } R\in (R_1;\, \infty) \\ 
\end{array} 
\right. 
\qquad  
V(R)= \left\{ 
\begin{array}{rcl}
\sqrt{\frac{f_0}{2 R_1}}\cdot R \makebox{\quad for  } R\in [0;\, R_1] \\ 
\sqrt{\frac{f_0\cdot R_1}{2}} \makebox{\quad for  } R\in (R_1;\, \infty) \\ 
\end{array} 
\right. 
\label{3-5}\end{eqnarray}
In this case the analogue of the integral (\ref{3-2}) has the form: 
\begin{eqnarray}
2\pi G\sigma (\rho)= \frac{2f_0}{\pi}\int\limits_{\alpha_2}^{\pi/2} 
\sqrt{1-(\rho\cos\alpha /R_1)^2} \, d\alpha \, ,  
\quad \alpha_2=\left\{ 
\begin{array}{rcl}
0 \makebox{\quad for  } \rho\le R_1 \\ 
\arccos(R_1/\rho) \makebox{\quad for  } \rho >R_1 \\ 
\end{array} 
\right.
\label{3-6}\end{eqnarray} 
This integral is expressed through the elliptic integral of the 2nd kind, 
here the asymptotic behaviour at ${\rho /R_1 \to 0}$ is nonsingular and we have that ${\sigma(\rho)\to const}$. 
At infinity, the asymptotic behaviour remains the same: That is, at ${\rho /R_1\to \infty}$ we have ${\quad\sigma(\rho)\propto (R_1/\rho)}$.

\section{OBSERVATIONAL DATA}
\label{s5} 

By using the method described above we have restored 
the matter surface density profiles for the discs of
4 spiral galaxies basing on their observed rotation curves. The four galaxies are :
a) NGC 2841, b) NGC 7217, c) NGC 7331, and d) NGC 5533
(the general characteristics of the galaxies are listed
in Table~\ref{tab1} and the restored surface density distributions are
presented in Fig.~\ref{R2}). 

We use for computing the two limiting cases (see Fig.~\ref{R2}) for the behavior of the curve ${V(R)}$ at large ${R>R_{max}}$ 
(where the velocity distribution is unknown): 

1. ${V=V(R_{max})=const}$

2. ${V=V(R_{max})\cdot\sqrt{R_{max}/R}}$ 

The second case corresponds to a Keplerian tail in the velocity distribution of matter, 
quantities corresponding to this case, marked on Fig.~\ref{R2} by a prime.

The galaxies under consideration are
all giant spiral galaxies of the Sab-Sb type and all of them
are classified as unbarred This is important because the axisymmetry
of the galaxies signifies regular circular rotation of the gas,
providing a possibility to measure their rotation curves from
a single spectral observation per galaxy if the spectrograph 
long slit is aligned with the disk isophote major axis. 
About accounting for the components of the spherical halo see section~\ref{s6}. 
The rotation curves have been obtained by measuring the line-of-sight velocities
of the gas. Because the gas is a cold dynamical subsystem and it is
confined to the thin disks, its rotation traces the gravitational
potential distribution very well. For the inner parts of the rotation curves we
have used spectral observations of the ionized (warm) gas which
have a good spatial resolution (typically about 1 arcec that corresponds
to a few dozen of parsecs for 3 of the 4 galaxies). For the outer parts, we used
radio observations of the neutral hydrogen at 21 cm which has a
spatial resolution that is worse by an order of magnitude compared to the optical.
However, the neutral hydrogen disks are usually more extended than the stellar
ones which allows us to trace the rotation curves (and thus the gravitational
potential distribution associated with them) much farther from the
galactic centres than where the stars are seen. 
The literature sources which we have used to construct the combined rotation curves for our 4 galaxies
are listed in Table~\ref{tab2}.

\begin{table}
\caption[ ] {Global parameters of the galaxies}
\begin{center}
\begin{tabular}{lcccc}
\hline\noalign{\smallskip}
NGC & 2841 & 5533 & 7217 & 7331  \\
\hline
Type (NED$^1$)\label{tab1} & SA(r)b & SA(rs)ab & (R)SA(r)ab & SA(s)b \\
$R_{25}$, arcsec (LEDA$^2$) & 244 & 93 & 117 & 314 \\
$R_{25}$, kpc  & 16.6 & 19.3 & 9.3 & 23 \\
Distance, Mpc (NED) &  14 & 43 & 16.5 & 15 \\
$M_B$ (LEDA)  & --20.84 & --21.6 & --20.48 & --21.56  \\
$V_{rot} \sin i$, $\mbox{km} \cdot \mbox{s}^{-1}$,
(LEDA, HI) & 295 & 206 & 143 & 237 \\
$M_{HI}$, $(10^9\,M_{\odot})^3$ & 2.8 & 30 & 0.7 & 8.2 \\
\hline
\multicolumn{5}{l}{$^1$\rule{0pt}{11pt}\footnotesize
NASA/IPAC Extragalactic Database}\\
\multicolumn{5}{l}{$^2$\rule{0pt}{11pt}\footnotesize
Lyon-Meudon Extragalactic Database}\\
\multicolumn{5}{l}{$^3$\rule{0pt}{11pt}\footnotesize
from \cite{Bosma1981} for NGC 2841 and NGC 7331 and from \cite{Noordermeer2005} 
for NGC 5533 and NGC 7217}
\end{tabular}
\end{center}
\end{table}

It is interesting to compare the restored density distributions of the
gravitating matter with the distributions of the stellar component in
the galaxies considered; in all of them the stars are the dominating baryonic content.
For this purpose we have used surface photometry data from the literature for
our galaxies. NGC 7331 and NGC 2841 have extended radial profiles
of the surface brightness in the NIR (2 mkm) photometric band $K$ published
in the survey by \cite{Munoz-Mateos2009}, the other two galaxies have been
studied by us earlier in the $R$-band (NGC 5533, see  \cite{Sil'chenko1998})
and in the $I$-band (NGC 7217, see \cite{Silchenko2000}). To derive a
density distribution from the brightness distribution or to calculate the
brightness distribution corresponding to some density distribution, we need
to know the mass-to-light ratio in the specified photometric band appropriate for
the stellar population of the particular age and metallicity. At the same time, the
latter characteristics are only known for the disks of a few nearby spiral
galaxies. For NGC 7217 and NGC 5533 we have studied the radial distributions
of the mean age and metallicity of the stellar populations in their disks
by analysing long-slit spectral data from the reducer SCORPIO of the Russian
6-meter telescope \cite{scorpio}. So for these galaxies we are able to use
the appropriate mass-to-light ratio values obtained in the evolutionary synthesis
models by \cite{Percival2009} (and re-calculated by us for the reduced Salpeter IMF so
slightly increased with respect to the models by \cite{Percival2009} which used the
Kroupa IMF).
For the other two galaxies, NGC 2841 and NGC 7331, the disk stellar population
parameters are unknown; for them we have used the empirical calibrations of the
mass-to-light ratio values versus broad-band colours from \cite{Bell2003} -- one versus
$g-i$ in NGC 2841 and the other versus $B-R$ in NGC 7331 (the broad-band colour profiles
for them are also taken from the photometric survey \cite{Munoz-Mateos2009}).

\begin{table}
\begin{center}
\caption[ ] {References for the rotation curves and photometric profiles}
\begin{tabular}{|c|c||c|c|} 
\hline\noalign{\smallskip}
NGC  &              References                      & NGC   &    References                              \\
\hline\hline 
2841 \label{tab2} & rot(H$\alpha$+[NII]): \cite{Afanasiev1999}   & 7217  & rot(H$\alpha$+[NII]): \cite{Zasov1997}     \\
     & rot(HI): \cite{Begeman1987, Begeman1991}     & \quad & rot(HI): \cite{Noordermeer2005}            \\
     & density(HI): \cite{Bosma1981}                & \quad & density(HI): \cite{Noordermeer2005}        \\
     & phot: \cite{Munoz-Mateos2009}                & \quad & phot: \cite{Noordermeer2007}       \\
\hline
5533 & rot(H$\alpha$+[NII]): \cite{Sil'chenko1998}  & 7331  & rot(H$\alpha$+[NII]): \cite{Afanasiev1989} \\
     & rot(HI): \cite{Noordermeer2005}              & \quad & rot(HI): \cite{Begeman1987, Begeman1991}   \\
     & density(HI): \cite{Broeils1994}              & \quad & density(HI): \cite{Bosma1981}              \\
     & phot: \cite{Sil'chenko1998, Noordermeer2007} & \quad & phot: \cite{Munoz-Mateos2009}              \\
\hline
\end{tabular}
\end{center}
\end{table}

\begin{figure*}
\includegraphics[width=15cm]{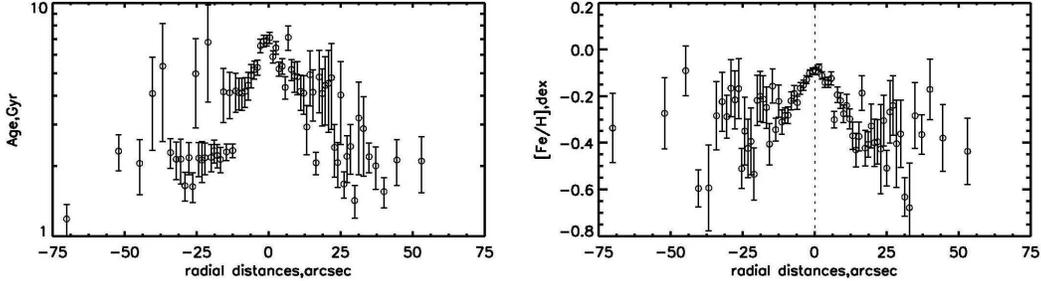}
\caption{Radial dependencies of the mean age and mean metallicity of the stellar populations in NGC 5533,
which we have obtained with the long-slit spectral data from the reducer SCORPIO of the Russian 6m telescope.}
\label{R4}
\end{figure*} 

Earlier we have found two large-scale stellar disks with different characteristics of
the stellar populations in NGC 7217
\cite{Sil'chenko2011-1, Sil'chenko2011-2}:  inside $R\approx 45''$ a thin stellar disk
with the mean stellar age of 5 Gyr and the mean metallicity of [Z/H]$=-0.2$ dominates.
At larger distances from the center a thicker and more extended star forming stellar disk 
with a younger, $\langle T \rangle =1-2$ Gyr, and very metal-poor, [Z/H]$\le -0.4$,
stellar population is seen. The border between two disks is rather narrow so a jump in
the mass-to-light ratio is noticeable at $R\approx 50''$; the matter density profile
(Fig.~\ref{R2}) demonstrates a break at the same radius. NGC 5533 is also a multicomponent
stellar system, however this time the limits of the subsystems do not coincide in the
brightness and density profiles: all the subsystems are more extended as presented by
the gravitating matter density distribution. The surface photometry allows to distinguish
a bulge-dominated region within $R<20''$, an inner stellar disk at $R=20''-60''$, and
an outer stellar disk of low surface brightness at $R>70''$ \cite{Sil'chenko1998}.
By applying the software ULYSS \cite{ulyss} to the long-slit spectral data
obtained with the SCORPIO of the 6m telescope, we have succeeded to derive
the stellar populations characteristics up to $R\approx 50''$,
so only for the inner stellar disk. The results are shown in Fig.~\ref{R4}. One can see 
that at $R>20''$, or beyond the bulge-dominated area, both the age and the metallicity
demonstrate flat radial dependencies. So for the inner stellar disk in NGC 5533 we can
use the mean values of the stellar population characteristics at $R>20''$, namely,
$\langle T \rangle =2$ Gyr and [Z/H]$=-0.4$. We have taken the model mass-to-light ratio
just for these stellar population parameters from \cite{Percival2009}, and then have added 0.15
in the logarithm to re-calculate it with the reduced Salpeter initial mass function (IMF)
instead of the Kroupa IMF used by \cite{Percival2009}
(the recommendation are taken from \cite{Bell2003}).

\begin{figure*}
\includegraphics[width=15cm]{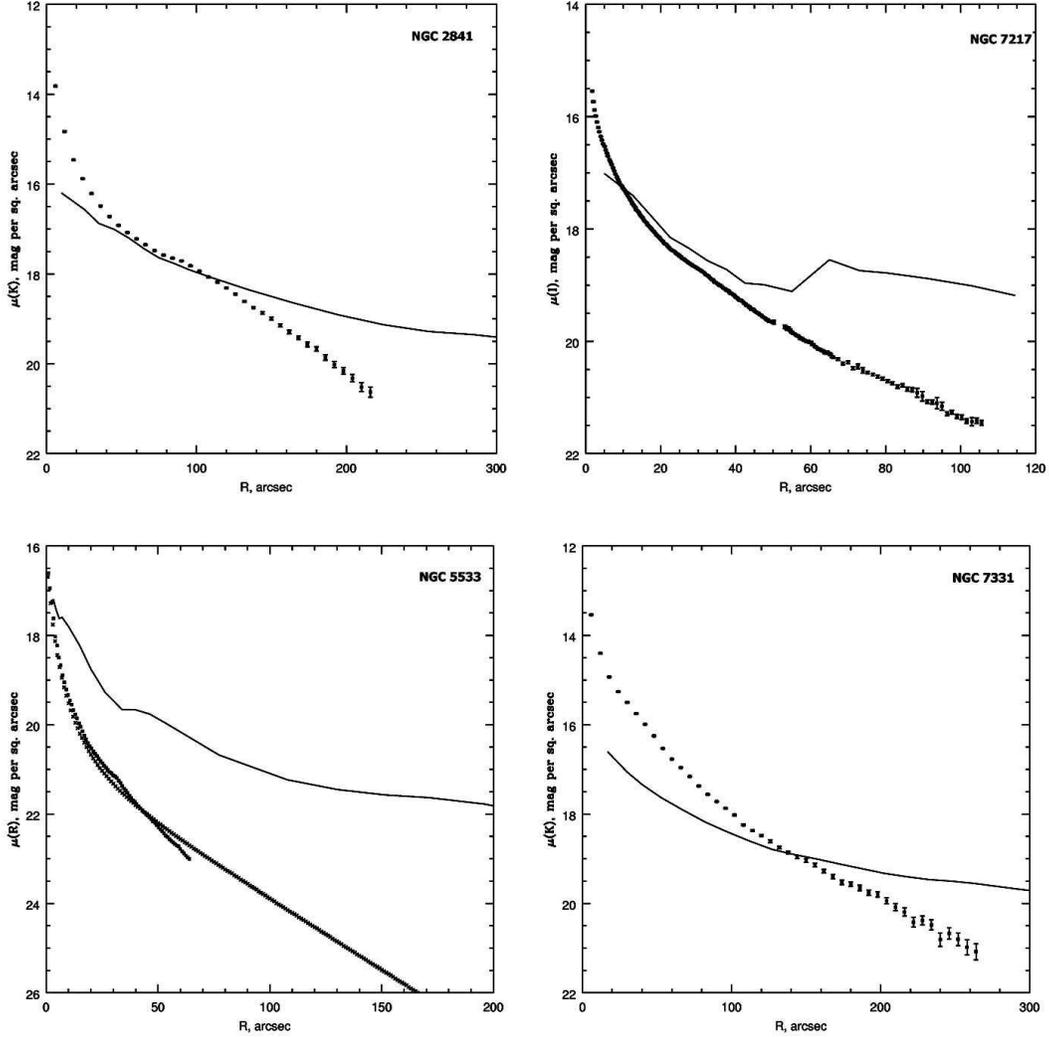}
\caption{The comparison of the observed surface brightness profiles for our four galaxies (signs), with the ones based on the gravitating matter surface density profiles restored from the rotation curves for the outer rotation curve asymptotic taken as a constant value (lines) calculated under the assumption that all the gravitating matter is the stellar component of the galactic disks.}
\label{R3}
\end{figure*} 

So in Fig.~\ref{R3} we confront the observed surface brightness profiles with those calculated from
the matter density profiles reconstructed from the rotation curves. If the main contributor into
the gravitational potential of a galaxy is a stellar disk, two profiles must coincide. What have 
we discovered from the Fig.~\ref{R3}?
We see that the different galaxies demonstrate different contributions of their stellar disks into
the total gravity. In NGC 2841 and NGC 7217 the inner parts of both profiles are in agreement;
in these galaxies, indeed, within $R<9$ kpc in NGC 2841 and within $R<4$ kpc in NGC 7217, the
stellar disks are the main contributors into the total gravitational potentials. In the outer
parts of the galaxies a `hidden mass' reveals itself -- the profiles reconstructed from the
rotation curves go strongly above the observed surface brightness profiles. But in other two
galaxies, NGC 5533 and NGC 7331, the observed and calculated profiles diverge everywhere in
the galaxies. In NGC 5533 the `hidden mass' dominates in the gravitational potential over the
whole galaxy extension -- it is just the same conclusion that we have already
made in our previous work basing on the analysis of the rotation curve shape \cite{Sil'chenko1998}.
NGC 7331 represents a more intriguing situation: the observed central brightness
exceeds our expectations arisen from the low visible rotation velocity. We think
that non-circular gas motions are significant in the center of this galaxy
so the rotation curve derived from the single long-slit cross-section along
the isophote major axis does not reflect a circular speed which characterizes
the gravitational potential. Our suspicion about the tri-axiality of the
gravitational potential shape in NGC 7331 has also been expressed by us
earlier \cite{Sil'chenko1999}.

\begin{figure*}
\includegraphics[width=15cm]{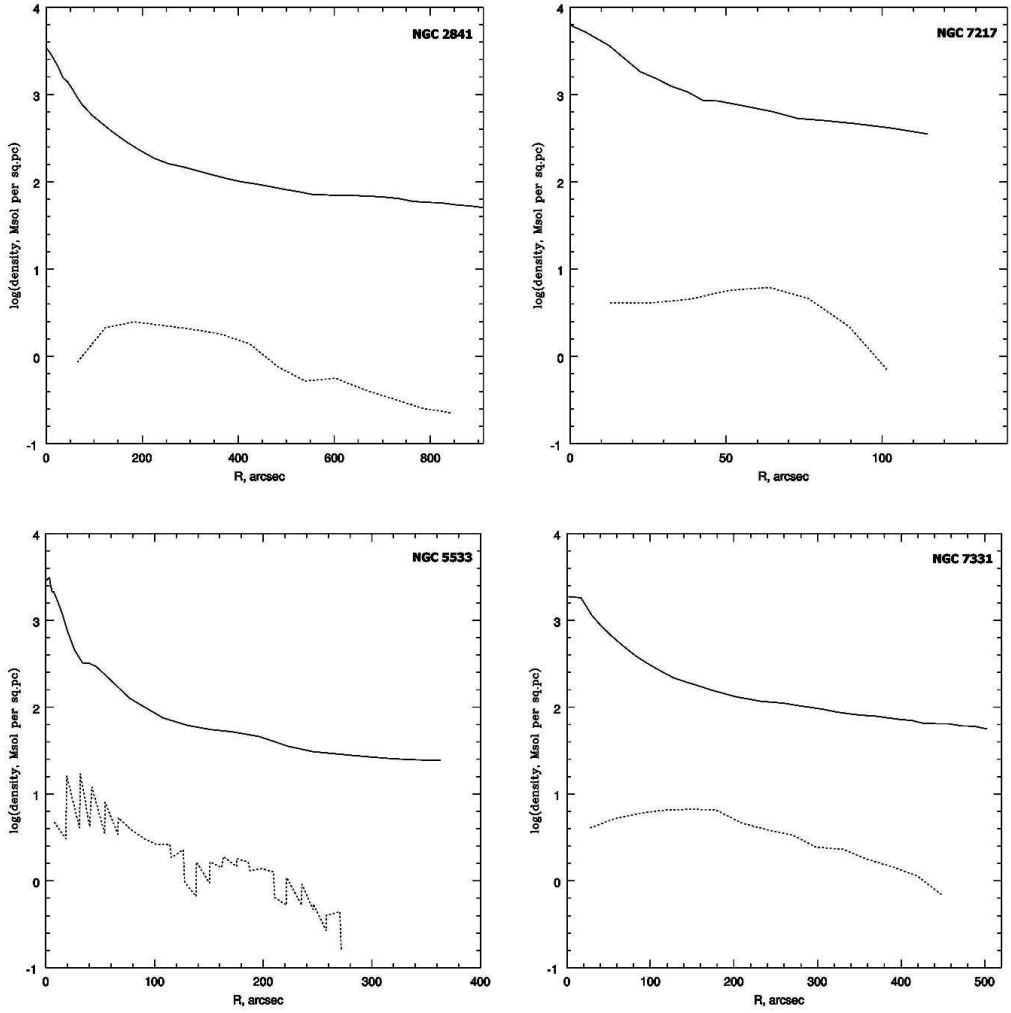}
\caption{The comparison of the gravitating matter density profiles reconstructed by us from the rotation curves with the outer rotation curve asymptotic taken as a constant value with the neutral hydrogen radial distributions in four galaxies under consideration.}
\label{R5}
\end{figure*}

What properties of NGC 2841 and NGC 7217 may be related to the dominance of their large-scale stellar
disks in their total gravity? We can do a couple of suggestions. First of all, both galaxies are
classified as isolated and are included into the list of isolated galaxies by \cite{Karachenceva1973}.
Secondly, there is a noticeable difference in neutral hydrogen content between NGC 2841 and NGC 7217,
on one hand, and NGC 5533 and NGC 7331, on the other hand: the latter pair of galaxies is more gas-rich
(see the Table~\ref{tab1}). In Fig.~\ref{R5} we compare radial profiles of the gravitating matter density  
reconstructed by us from the rotation curves, to the neutral hydrogen surface density radial distributions
(at the logarithmic scales) for all four galaxies. While in NGC 2841 and NGC 7217 the neutral hydrogen
contributes less than one percent of the total surface density, in NGC 5533 and NGC 7331 its contribution
is as high as a few percent over the whole extension of the galactic disks. Moreover, the total character
of the gas density and matter density profiles is the same: they look like a quasi-exponential decreasing
function which is very typical just for stellar and gaseous disks of galaxies. The remark that the
`hidden mass' distribution in spiral galaxies is quite similar to the `visible' gas distribution has been
already made earlier by e.g. \cite{Hoekstra2001}. The fact found by us in this work that the gravitating 
matter surface density distribution agrees better with that of the stellar component in the gas-poor galaxies
is therefore in line with the hypothesis by \cite{Pfenniger1994} that the `hidden mass' in spiral galaxies
is in fact very cold molecular gas which cannot be observed in accessible electromagnetic radiation spectral 
domains.

\section{Accounting for the components of a spherical halo of the galaxy}
\label{s6} 

In all previous sections we ignored the fact that much of the matter of the galaxy, in principle, can be concentrated in a spherically symmetric halo, which has a mass function ${M_{sphere}(r)}$. 
Initially, we do not know the distribution of the spherical mass ${M_{sphere}(r)}$, this feature is a free parameter model. 

Due to the additivity of the potential $\Phi$ (in the Newtonian approximation), it can be divided into two parts (the disk and a spherical component): 
\begin{eqnarray}
\Phi (\rho ,z)=\Phi_{disc} (\rho ,z) + \Phi_{sphere} (r) \, ,\quad r^2=\rho^2+z^2 . 
\label{6-1}\end{eqnarray} 
Then equation (\ref{1-1}) in the equatorial plane ($ {z = 0} $) can rewritten in as\footnote{Here we took into account that for the spherical part of potential we have: ${\Phi_{sphere}(r),_r = GM_{sphere}(r)/r^2}$.}: 
\begin{eqnarray}
\frac{\partial\Phi_{dics}(\rho , z=0)}{\partial\rho} = \frac{V^2(\rho) - GM_{sphere}(\rho)/\rho}{\rho} 
\label{6-2}\end{eqnarray} 
Hence, all subsequent arguments (starting with the formula (\ref{1-1})) amounts to replacing the observed features ${V(\rho)}$ as: 
\begin{eqnarray}
V(\rho)\to\sqrt{V^2(\rho) - GM_{sphere}(\rho)/\rho} 
\label{6-3}\end{eqnarray} 
Here we are free to vary the expression ${GM_ {sphere} (\rho) /\rho}$ in the range from zero to a value approximately equal to  
${V^2(\rho)-GM_{disc}(\rho)/\rho}$, where ${M_{disc}(\rho)}$ denotes the mass which is concentrated in the disk to a certain radius $\rho$. 
You can, for example, from a certain radius (e.g. the radius of the observed stellar disk to the end of the 25-th blue isophotes), put ${V^2(\rho) - GM_{sphere}(\rho)/\rho = GM^{tot}_{disc}/\rho}$, where $M^{tot}_{disc}$ is the total mass of matter in the galactic disk. 
This will correspond to the termination of the matter density in the disk at this radius (25-th blue isophotes).

\section{Conclusions} 
\label{conclusion}

In this paper we considered a new method for reconstructing of the surface
density of matter in a flat disk of spiral galaxies. 
At the beginning we suppose that all matter is concentrated in the disk of the
galaxy and apply the method for reconstructing the surface density of matter in
4 spiral galaxies: NGC2841, NGC7217, NGC7331 and NGC5533. 
We compared our results with the distributions of the stellar matter. 
This comparison allows us to obtain some conclusions about dark matter in these
galaxies. 
In section~\ref{s6} we discuss possible accounting for the components of the
spherical halos.

\section*{ACKNOWLEDGEMENTS}
\label{ask}

We thank Bratek L., Jalocha J., Kutschera M., Skindzier P. and Polyachenko E.V. for valuable comments and discussions of this work. 

This work was supported in part by the Federal Programm "Scientific and Pedagogical Personal of Innovative Russia 2009-2013", 
and by the program of the Presidium of Academy of Sciences "Origin, structure and evolution of objects in the Universe 2011".    
Jakob Hansen was supported in part by the Research Grant from SUN Micro Systems at KISTI.

\appendix

\section{Numerical Integration of our new method}
\label{appA}

As described in the main text, we wish to use our new method for finding
distributions of surface density and mass throughout a galactic disk, i.e. eqs.
\ref{2-9} and \ref{2-10}. Due to the observational nature of our input data (the
velocity function), our only means for solving these equations are by numerical
methods. This appendix describes and tests our numerical code developed for this
purpose.

The fundamental integral of our method, eq. \ref{2-9}, do not have
non-integrable singularities, however, the integrand may be ill-defined at the
integration endpoints. Because of that, we use the open-ended second
Euler-Maclaurin integration formula as our basic integration algorithm.
Secondly, we use increasing numerical resolutions and Romberg's method to
achieve the highest possible accuracy (see discussion on these techniques
and further references e.g. in \cite{Pre92}). We use these methods with
enough iterations until the result has satisfactorily converged.

In addition to these basic integration techniques, a few other "tricks" are
needed. Our basic data for integration will usually be a discrete set of
velocity data in the disk as a function of the radius, $\rho$. Obviously, eq.
\ref{2-9} is a function of the velocity function, but because the velocity
function is expressed through the intermediate variable $R_0 = \rho \frac{\cos
\alpha}{\sin \beta}$ (see eqs. \ref{2-8}-\ref{2-9}), we will often need values of
the velocity function not explicitely tabulated by our observational data. To
solve this issue we need to use interpolation for velocity data in between
observational data points and also we need to make some assumptions about the
behaviour of the velocity function at points beyond the tabulated range of
observational data. 

To obtain velocity data for points in between the observational data points, we
have used simple linear interpolation (attempts with using higher order
interpolations were made, but yielded no increased accuracy). Furthermore, we
have made the assumptions that 1) the velocity at the central axis is zero and
2) that the velocity beyond the outermost observed point is a constant function
with a value equal to that of the outermost observed point. 

\begin{figure*}
\includegraphics[width=14cm]{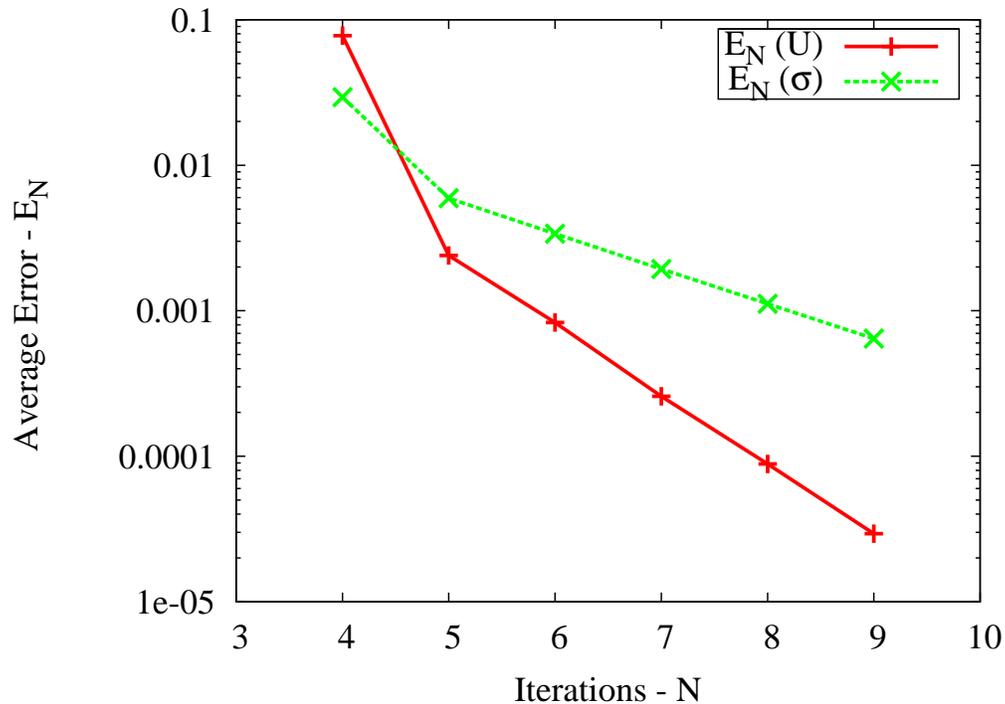} 
\caption{{Demonstration of convergence of basic functions of our method.}}
\label{app-fig1}
\end{figure*}

Finally, we have tested our numerical implementation of the new method by doing
a number of convergence tests. We define an average error measure as :
\begin{eqnarray}
E_N (x) = \frac{1}{n} \sum_{i=1}^n \frac{|x_N^i - x_{N+1}^i|}{x_{N+1}^i},
\label{app-eq1}\end{eqnarray} 
where $x$ denotes the function we are investigating, $n$ denotes the number of data points (in $\rho$) and $N$ 
denotes the number of iterations we use in Romberg's method\footnote{I.e. "$N+1$" indicates higher resolution than "$N$".}. 
The data in figure \ref{app-fig1} uses this error measure for the two basic basic functions $U( \rho )$ and $\sigma (\rho )$
(c.f. eq. \ref{2-9}) for varying number of iterations. These two functions are the most basic functions of our new method and
if the method is correctly numerically implemented, the average error must converge to zero as we increase the numerical resolution (number of 
Romberg iterations).

As the input data for this figure we have used the test velocity function 
eq. \ref{velocity-test-func}, but it is noted that similar results are obtained if actual observational velocity
data is used. 
From the figure it is clearly seen that both $U( \rho)$ and $\sigma (\rho )$ are converging towards zero with incresing number of iterations. 
This is exactly what we expect and we thus conclude that the numerical implementation of our new method is correct.

Finally it is noted that for all numerical results presented in this paper, convergence tests have been made to
confirm that the presented data has converged to its limiting value.

\bibliography{mybib}

\end{document}